\documentclass[aps,prx,twocolumn,showpacs,superscriptaddress,groupedaddress]{revtex4-2}
\usepackage{graphicx}
\usepackage{dcolumn}
\usepackage{bm}

\usepackage{amssymb}
\usepackage{epsfig}
\usepackage{amsmath}
\usepackage{subfigure}
\usepackage{textcomp}
\usepackage{url}
\usepackage{float}
\usepackage{color}
\usepackage{cases}
\usepackage[normalem]{ulem}

\usepackage[colorlinks=true, urlcolor=blue, anchorcolor=blue, citecolor=blue,filecolor=blue,linkcolor=blue,menucolor=blue
]{hyperref}

\newcommand{\OO}{\mathcal{O}}

\newcommand{\psiu}{\psi_{+}}
\newcommand{\psid}{\psi_{-}}
\newcommand{\phiu}{\phi_{+}}
\newcommand{\phid}{\phi_{-}}
\newcommand{\pd}[2]{\frac{\partial #1}{\partial #2}}

\newcommand{\tasd}{\overline{\delta^2(\Delta)}}

\newcommand{\average}[1]{\left< #1 \right>}
\newcommand{\Pnml}[1]{P_{n, m}^{(\ell)}(#1)}

\begin{document}

\preprint{APS/123-QED}

\title{Non-Markovian gene expression}

\author{Ohad Vilk $^{a,b,c}$}
\author{Ralf Metzler$^{d,e}$}
\author{Michael Assaf $^{a}$}
 \affiliation{$^a$ Racah Institute of Physics, The Hebrew University of Jerusalem, Jerusalem 91904, Israel,}
 \affiliation{$^b$ Movement Ecology Lab, Department of Ecology, Evolution and Behavior, Alexander Silberman Institute of Life Sciences, Faculty of Science, The Hebrew University of Jerusalem, Jerusalem 91904, Israel,}
 \affiliation{$^c$ Minerva Center for Movement Ecology, The Hebrew University of Jerusalem, Jerusalem 91904, Israel}
 \affiliation{$^{d}$Institute of Physics and Astronomy, University of Potsdam, Potsdam 14476, Germany}
 \affiliation{$^{e}$Asia Pacific Centre for Theoretical Physics, Pohang 37673, Republic of Korea}
 

\begin{abstract}
    We study two non-Markovian gene-expression models in which protein production is a stochastic process with a fat-tailed non-exponential waiting time distribution (WTD). For both models, we find two distinct scaling regimes separated by an exponentially long time, proportional to the mean first passage time (MFPT) to a ground state (with zero proteins) of the dynamics, from which the system can only exit via a non-exponential reaction. At times shorter than the MFPT the dynamics are stationary and ergodic, entailing similarity across different realizations of the same process, with an increased Fano factor of the protein distribution, even when the WTD has a finite cutoff. Notably, at times longer than the MFPT the dynamics are nonstationary and nonergodic, entailing significant variability across different realizations. The MFPT to the ground state is shown to directly affect the average population sizes and we postulate that the transition to  nonergodicity is universal in such non-Markovian models. 
\end{abstract}

\maketitle

\textit{Introduction.} Gene expression is an inherently stochastic process, playing a key role in the function of  prokaryotes and eukaryotes~\cite{elowitz2002stochastic, blake2003noise, kaern2005stochasticity, shahrezaei2008analytical,bar2006noise}. Stochastic gene expression in genetically identical cells in identical environments is governed by mRNA and protein noise, and is suggested to be an important source of phenotype variability~\cite{raj2008nature, raser2005noise, munsky2012using}, and an important trait that can optimize the balance of fidelity and diversity in eukaryotic gene expression~\cite{raser2004control}.

Multiple studies have treated stochastic gene expression using either chemical master equations or Langevin equations in various models~\cite{ackers1982quantitative,gardiner1985handbook,   hornos2005self,shahrezaei2008analytical, morelli2008reaction,
assaf2011determining,
assaf2013extrinsic,  roberts2015dynamics, yin2021optimal}. 
However, most existing models are Markovian with exponentially distributed inter-reaction times~\cite{gardiner1985handbook}. Although this is a valid assumption in many realistic cases~\cite{aurell2002epigenetics}, it is becoming apparent that many natural processes exhibit long delays or non-exponential intrinsic waiting times~\cite{golding2023gene,  jo2014analytically, zhang2021analysis}. Molecular memory can be created, e.g., due to incomplete mixing of small reaction steps involved in the synthesis of macromolecules, such as mRNA or proteins~\cite{bratsun2005delay, brett2013stochastic, schwabe2012transcription, jia2011intrinsic, baghram2019exact}. 
Moreover, even in simple geometries, reaction dynamics are characterized by an enormous spread of relevant time scales~\cite{godec2016universal, 
grebenkov2018strong, grebenkov2020single}, such that reaction and diffusion control are intricately coupled, in contrast to models based on (global) mean first passage and reaction times~\cite{collins1949diffusion, coppey2004kinetics, condamin2007first, benichou2010geometry}. Such defocused reaction times are relevant for intra-cellular regulation for low-concentration reactants. 
In particular, processes governed by power-law waiting times were shown to determine the motion of protein channels in membranes of living cells, displaying diffusion-controlled anomalous dynamics~\cite{weigel2011ergodic}. 

To study processes with fat-tailed (e.g., power-law) WTDs between steps, displaying aging behavior and ergodicity breaking~\cite{bel2005weak, burov2010aging, barkai2012single, metzler2014anomalous,weigel2011ergodic, jeon2011vivo,tabei2013intracellular, FoxPowerSpectrum, vilk2022ergodicity, vilk2022unravelling}, one often uses the continuous-time random walk (CTRW) framework~\cite{metzler2000random, metzler2014anomalous}. Recently, a chemical CTRW master equation has been  suggested to analyze non-Markovian birth-death dynamics~\cite{aquino2017chemical,leier2015delay,yin2021optimal, zhang2021analysis, jiang2021neural}. 
Yet, a systematic study on gene expression with intrinsic reactions possessing a fat-tailed WTD (with a finite or infinite mean), has not yet been carried out. 

Here, we apply the CTRW formalism to gene expression models with intrinsic fat-tailed WTD   depending on the system's current state, and identify a general class of  models transitioning between ergodic and nonergodic phases at non-trivial long times, when  the WTD's mean diverges. In addition, stationarity and ergodicity are shown to strongly depend on the system's internal noise.

\begin{figure}[t]
    \includegraphics[width=0.39\textwidth,clip=]{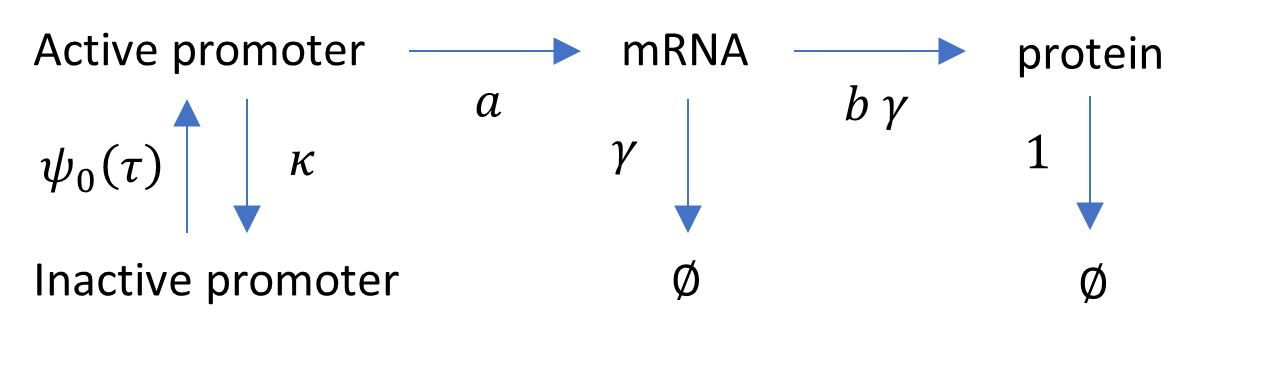}
      \vspace{-4mm}  \caption{A two-state gene expression model with a promoter, transcription, and translation. The activation process has a fat-tailed WTD denoted by $\psi_0(\tau)$. All other processes have exponential WTDs, and their average rates are specified.}
    \label{fig:fig1}
    \vspace{-4mm}
\end{figure}

We consider a two-state model in which a promoter randomly transitions between transcriptionally active and inactive states, see Fig.~\ref{fig:fig1}. In contrast to previous models, we assume that the time it takes the promoter to activate is power-law distributed, mimicking delays due to the DNA binding to specific and limited elements in the cell. This may be justified by realizing that activating the promoter often requires binding to sparse elements in the cell~\cite{coppey2004kinetics, kurilovich2020complex, doerries2022rate}. Thus, the activation WTD is related to mean time it takes a random walker to hit a target, 
which follows a power-law distribution~\cite{gardiner1985handbook, bray2013persistence}.

To gain insight into the two-state model depicted in Fig.~\ref{fig:fig1}, we first study a simpler model of a self-regulating gene (SRG) with linear rates. Here, the WTD between protein production events is assumed to be fat-tailed, as protein production requires the DNA state to be active. We show that even this simple model is sufficient to give rise to non-trivial dynamics, including the existence of a long-lived metastable state followed by protein decay. To this end, we compute the typical time to transition from  metastability to power-law decay, the mean protein number, and its copy-number distribution. The latter displays super-Poissonian behavior, with a Fano factor greater than 1 (see below), even for truncated WTDs, which are experimentally relevant~\cite{chen2015genome}. Finally, we  study the full two-state model (see Fig.~\ref{fig:fig1}) and show that the two models are qualitatively identical. Our theory is verified via numerical simulations based on a modified Monte-Carlo (MC) algorithm, recently developed for non-Markovian stochastic systems~\cite{anderson2007modified, boguna2014simulating,masuda2018gillespie}.

\textit{Self-regulating gene}. We consider a protein-only model for the number of proteins $n$ defined by the  reactions $ n \rightarrow n+1$ and $n \rightarrow  n-1 $, 
respectively representing protein production, e.g., due to translation, and  protein degradation, e.g., due to cell division. Most studies assume that these reactions are exponential,  memory-less point processes~\cite{gardiner1985handbook}. Yet, in realistic scenarios, the times between consecutive reactions is not necessarily exponentially distributed~\cite{golding2023gene} and the process may be described by a WTD of the next event. We thus define $\psid(\tau)$ as the WTD for one of $n$ proteins to degrade between  $t$ and $t+\tau$ ($n \rightarrow n-1$), and $\psiu(\tau)$ as the WTD for protein production ($n \rightarrow n+1$). These are given by 
\begin{equation} \label{psi1}
    \psid(n, \tau) = n e^{-n \tau }, \; \psiu(\tau) = K / [1 + K \tau/\alpha]^{1+\alpha}. 
\end{equation} 
Here, degradation is assumed to be a Poisson process linear in the population size, which gives rise 
to an exponential WTD of $\psid(\tau)$~\cite{gardiner1985handbook} with normalized rate 1. 
In contrast, as discussed above, production may depend on a larger number of intra-cellular products, present in small numbers~\cite{hornos2005self, kaern2005stochasticity} or may take place in a sparse environment, leading to fat-tailed delays. Thus,  production events are no longer exponentially distributed but sampled from a fat-tailed distribution. 
For concreteness, we consider the power-law WTD in \eqref{psi1}. Here $K$ is the carrying capacity and $\alpha$ is the power-law exponent.


To derive the birth-death master equation we define the probability  $\phiu(n, t)$ [$\phid(n, t)$] for a single protein production  (degradation) event to occur at time $t$ provided that no degradation (production) event occurred until time $t$, when the system has $n$ proteins, which reads
\begin{equation} \label{phi2}
    \phi_i(n, t) = \psi_i(t) \int_t^\infty \psi_j(\tau) d\tau, 
\end{equation}
for $i, j \in \{+, -\}$ and $i\neq j$. 
Using the CTRW formalism developed in Ref.~\cite{aquino2017chemical} we write the following chemical master equation for the probability $P_n(t)$ of having $n$ proteins at time $t$, given Eqs.~\eqref{phi2}: 
\begin{equation} \label{MasterEquation}
    \frac{dP_n}{dt}\! =\! (E_n^1 \!-\!1) n P_n(t)  + (E_n^{-\!1} \!-\!1)\!\!\!\int_0^t\!\! \!M(n, t\!-\!t') P_{n}(t') dt'\!, 
\end{equation}
where  $E_k^j f(k) = f(k + j)$ are step operators.
The kernel $M(n, t)$ is defined in terms of its Laplace transform~\cite{aquino2017chemical}
\begin{equation} \label{memorykernel}
    \tilde{M}(n, s) = s \tilde{\phi}_+(n, s) /\left[1 - \tilde{\phi}_+(n, s) - \tilde{\phi}_-(n, s)\right], 
\end{equation}
where $\tilde{\phi}_i$ are the Laplace transforms of $\phi_i$ and $s$ is the Laplace variable. 
The term $(E_n^1 \!-\!1) n P_n(t)$ in Eq.~\eqref{MasterEquation} is due to the exponential degradation of $n$ products, while the integral term comes from the non-exponential production terms with memory kernel $M(n, t)$. Note that, in the case of an exponential WTD [i.e., if $\psiu(\tau) = K e^{-K \tau }$] one obtains $\tilde{\phi}_- = n/(s +K +n)$ and $\tilde{\phi}_+ = K/(s +K+n)$. Substituting these into Eq.~\eqref{memorykernel} yields $\tilde{M}(n, s) = n$ and $M(n, t) = n \delta(t)$. Thus, Eq.~\eqref{MasterEquation} reduces to the well-known chemical master equation for exponential WTDs~\cite{gardiner1985handbook}. 

\begin{figure}[t]
    \includegraphics[width=0.47\textwidth,clip=]{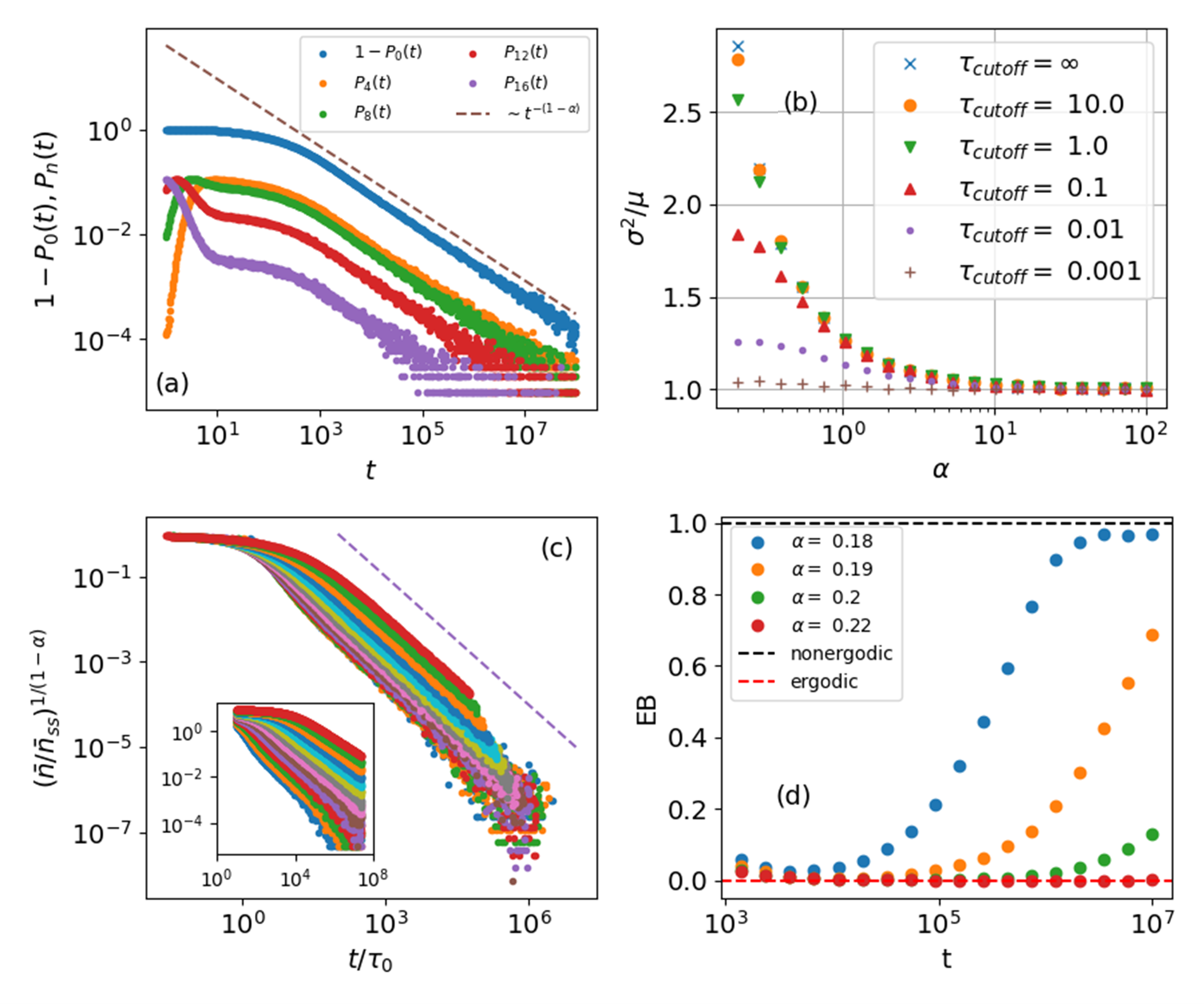}
        \vspace{-5mm}\caption{SRG model. (a) Probabilities $P_n(t)$ for various $n$ (see legend) and $1 - P_0(t)$ based on MC simulations for $K = 25$ and $\alpha = 0.36$, all showing similar scaling at long times. (b) The Fano factor versus $\alpha$ for different $\tau_{\text{cutoff}}$ (see legend), for $K = 500$.  (c) The mean protein number versus time for different values of $\alpha$ ($0.2-0.46$); the curves can be approximately collapsed by properly rescaling $\bar{n}$ and $t$, see text. The dashed line is an eye guide with a  slope of $-1$. Inset displays the averages without rescaling. Here $\bar{n}$ is averaged over $10^5$ simulations and the carrying capacity is $K = 25$. (d) The EB parameter versus time for different $\alpha$ values, $K = 500$ and $\Delta = 100$, see text.  }
    \label{fig:fig4}
    \vspace{-3mm}
\end{figure}

For a power-law WTD, the memory kernel is obtained by substituting Eqs.~\eqref{psi1} and \eqref{phi2} into Eq.~\eqref{memorykernel}: 
\begin{equation} \label{memorykernelexplicit}
\hspace{-2.5mm}\tilde{M}(n, s) = K  E_{\alpha +1}\left[\alpha(n+s)/K \right] /E_{\alpha}\left[\alpha(n + s)/K\right]\!,
\end{equation}
where $E_m(z) \equiv \int_1^{\infty}e^{-zt}t^{-m} dt$ is the exponential integral. 
Being interested in the long-time dynamics, $t \gg 1$, we approximate the memory kernel at $s \ll 1$, where states with $n > 0$ and $n=0$ (ground state) display a markedly different behavior. For $n > 0$ and $\alpha >0$, Eq.~\eqref{memorykernelexplicit} can be approximated as 
  $  \tilde{M}(n>0, s\ll 1) =  K m(x) + \OO(s), $
with $x \equiv n/K$ being the protein density, and 
$
    m(x) \equiv E_{\alpha +1}\left(\alpha x\right)/E_{\alpha}\left(\alpha x\right)
$. 
The leading-order inverse Laplace transform reads  $M(n>0, t) \simeq \delta(t) K  m(x)$; i.e., short-term memory for $n>0$, as degradation can always occur. 
In contrast, we find for $n=0$ and $s\ll 1$:
\begin{equation} \label{memorykernel_nzero_series}
    \hspace{-2.5mm}\tilde{M}(0, s) \!=\! K\!\times\!\begin{cases}
          \frac{(\alpha -1) }{\alpha }\left\{1 \!+\! \OO[\frac{\alpha s}{K},(\frac{\alpha s}{K})^{\alpha - 1}]\right\} & \alpha > 1, \\ 
         \frac{  \left(\alpha  s/K \right)^{1-\alpha }}{\alpha  \Gamma (1-\alpha )}\left\{1 \!+\! \OO[(\frac{\alpha s}{K})^{1-\alpha}]\right\}  & \alpha < 1      .
    \end{cases}
\end{equation}
Here, for $\alpha > 1$ the leading-order term is independent on $s$. In contrast, for $\alpha <1$ there is a power-law dependence, since at $n=0$ the only reaction that can occur is production, which gives rise to increasingly long times of inactivity.  Note that, the subleading term in Eq.~(\ref{memorykernel_nzero_series}) differs between $\alpha >2$ and $1<\alpha<2$, which affects the \textit{rate} of convergence to stationarity, but not stationarity itself. The inverse Laplace transform of \eqref{memorykernel_nzero_series} reads
\begin{equation} \label{memorykernel_nzero_series_time}
    M(0, t) \simeq K\times\begin{cases}
          \frac{(\alpha -1) }{\alpha } & \alpha > 1 ,\\ 
         \frac{(\alpha -1)K^\alpha \sin (\pi\alpha)}{\alpha^\alpha \pi t^{2-\alpha}}  & \alpha < 1 .    
    \end{cases}
\end{equation}
For  $\alpha > 1$, the dynamics is unaffected by the state $n=0$, as its probability, $P_0$, is exponentially small for $K\gg 1$. 
On the other hand, for $\alpha < 1$ the dynamics display long-range correlation even at infinitely long times. This is a clear signature of non-ergodicity. 
To show this, we derive an equation for the mean protein number $\bar{n}$ by multiplying Eq.~\eqref{MasterEquation} by $n$ and summing over all $n$:
\begin{equation} \label{nbar_protien_only}
    \pd{\overline{n}}{t} = - \overline{n}(t) + \int_0^t \overline{M(n, t-t')}dt', 
\end{equation}
where $\overline{M(n, t - t')} \equiv \sum_{n=0}^{\infty} M(n, t-t')P_n(t')$. This equation cannot be solved explicitly, and below we deal with it asymptotically. For $\alpha > 1$ we show below that at $t\gg 1$, $M(n, t) \to M(n) \equiv K m(x)$ and $\overline{m(x)}\simeq m(\overline{x})$, suggesting that the dynamics are ergodic. On the other hand, for $\alpha <1$, a single state retains memory leading to ergodicity breaking and aging~\cite{aquino2017chemical, metzler2014anomalous}. We conjecture that at sufficiently long times (see below),  $P_0(t)$ grows due to long periods of inactivity at $n=0$, whereas $P_{n>0}(t)$ rapidly decay. In Fig.~\ref{fig:fig4}(a) we show that for $t \gg 1$: $1 - P_0(t) \sim t^{-(1-\alpha)}$ and $P_{n>0}(t) \sim t^{-(1-\alpha)}$. The averaged memory kernel is then dominated by the state $n=0$ such that for $1 \ll t - t' \ll t$: $\overline{M(n, t-t')} \simeq \sum_{n=1}^{\infty} M(n)P_n(t') + M(0, t-t')P_0(t')  \sim M(0, t - t') \sim (t-t')^{-(2-\alpha)}$, where we have used~(\ref{memorykernel_nzero_series_time}) and that $P_0(t')\simeq 1$. Substituting this into~\eqref{nbar_protien_only}, the long-time asymptotic of the dynamics  for any $\alpha <1$ becomes: $\overline{n} \sim t^{-(1-\alpha)}$ at $t\to\infty$.

At what times is this scaling reached? As the scaling is caused by the state $n = 0$, the dynamics are effectively stationary as long as this state has not been visited. Only upon visiting $n=0$, we expect long-memory effects and nonstationarity. Thus, the typical time to asymptotic decay, $\tau_{typ}$, is  roughly proportional to $\tau_0$ -- the mean first passage time (MFPT), to reach  $n = 0$. At times $t\ll  \tau_{typ}$ the dynamics is expected to be stationary, allowing us to analytically find the long-lived metastable state prior to the asymptotic decay as well as the MFPT to $n = 0$. 

\textit{Stationary dynamics.} For $\alpha > 1$ and $t\gg 1$, and for $\alpha < 1$ and $1\ll t\ll \tau_{typ}$, the dynamics are stationary. To find $\bar{n}(t)$ and the MFPT to state $n = 0$,
we write a stationary master equation ($dP_n(t)/dt = 0$) by Laplace-transforming Eq.~\eqref{MasterEquation}, multiplying by $s$, and using the final value theorem: 
$\lim_{s\to 0} s \tilde{P}_n(s) = P_n $~\cite{yin2021optimal}. This yields: 
\begin{equation} \label{SteadyMasterEquation}
    0 = (n+1) P_{n+1} - n P_n + M(n-1) P_{n-1} - M(n) P_{n}, 
\end{equation}
where $P_n$ is the steady state solution for the probability to have $n$ proteins at time $t$, and $M(n) = K m(x)$, see definition below Eq.~\eqref{memorykernelexplicit}. 
Equation~\eqref{SteadyMasterEquation} can be solved recursively to give $P_n =  (1/n!)P_0\prod _{k=0}^{n-1} M(k)$, where $P_0$ is the probability to be in the ground state, found by normalization. For $K\gg 1$, this can be recast (up to a prefactor) into a semi-classical form as $P_n \sim e^{-K S(n/K)} \equiv e^{-K S(x)}$, with the action  
    $S(x) = \int_{\bar{x}}^x \ln \left( x'/m(x') \right) dx'$~\cite{dykman1994large,assaf2010extinction,assaf2017wkb}, where $\bar{x} \equiv \bar{n}/K$.
The integral here can be solved numerically for any $n$, 
and the MFPT to the ground state $n=0$ is given by $\tau_{0} \sim e^{K S(0)}$~\cite{dykman1994large,assaf2010extinction,assaf2017wkb}. 

In addition to the steady-state dynamics, in Fig.~\ref{fig:fig4}(b) we show that the Fano factor, defined as the variance of the protein number divided by its average ($\sigma^2/\mu$), increases with decreasing $\alpha$. As in all experiments the power-law WTD is expected to have an exponential cutoff at some finite $\tau_{\text{cutoff}}$, see e.g.,~\cite{chen2015genome}, we plot in Fig.~\ref{fig:fig4}(b) the Fano factor for different values of $\tau_{\text{cutoff}}$, see Appendix~\ref{appendixA}. Here the Fano factor is obtained from MC simulations and does not depend on the carrying capacity. Notably, the distribution tends to a Poissonian ($\sigma^2/\mu = 1$) either at $\alpha\gg 1$ or for very short cutoff times, $\tau_{\text{cutoff}} \lesssim 1/K$. 
The existence of reactions with fat-tailed WTDs can thus serve as a possible explanation of experimental observations of super-Poissonian distributions (with $\sigma^2/\mu>1$) in gene expression
~\cite{thattai2001intrinsic, choi2008stochastic}. 

In Fig.~\ref{fig:fig4}(c) we test our analytical results using simulations for a wide range of $\alpha <1$ values. To collapse the curves, and to show the asymptotic behavior of $\bar{n}\sim t^{-(1-\alpha)}$, we plot $(n/\bar{n}_{ss})^{1/(1-\alpha)}$, versus normalized time $t/\tau_0$, where $\bar{n}_{ss}$ is the numerical solution of~(\ref{nbar_protien_only}), 
$\tau_{0} = e^{KS(0)}$, and $S(0)$ is found numerically. 
The collapse indicates that all curves start decaying at roughly the same normalized time. Yet, the curves clearly do not perfectly overlap at long times, due to an intermediate regime in Fig.~\ref{fig:fig4}(c), caused by an $\alpha$-dependent prefactor, not accounted for by our theory; we have checked that the width  decreases as $K$ increases.
Finally, Fig.~\ref{fig:fig4}(d) shows further evidence of the crossover between ergodic and nonergodic dynamics at $\alpha$-dependent times, using the ergodicity breaking (EB) parameter~\cite{metzler2014anomalous}. 
The latter is the variance of the time-averaged squared-displacement of the protein number $\tasd \!=\! 1/(t\!-\!\Delta)\int_0^{t \!-\! \Delta} [n(t' \!+\! \Delta)\! -\! n(t')]^2 dt'$, divided by its mean, and is used as a measure for the level of variability across different trajectories of a given ensemble (see Appendix~\ref{appendixB} for additional details). 
The EB parameter is plotted in Fig.~\ref{fig:fig4}(d) versus the total simulation time, with EB$\ll\! 1$ and EB=${\cal O}(1)$, being signatures of ergodic and nonergodic dynamics, respectively~\cite{he2008random, metzler2014anomalous,thiel2014weak}.  

\textit{Two-state promoter model}. Similar effects occur in a more complex two-state gene expression model, which explicitly accounts for mRNA noise,  where transitions between a transcriptionally active and inactive promoter are independent of the protein number~\cite{shahrezaei2008analytical,assaf2011determining}, see Fig.~\ref{fig:fig1}. As stated above, activation often requires binding to limited  elements in the cell, which may give rise to a non-exponential, fat-tailed WTD [see Eq.~\eqref{psi1}]:
\begin{equation} \label{psi2_full}
\vspace{-1.0mm}
    \psi_0(\tau) = \kappa /  [1+\kappa \tau/\alpha]^{1+\alpha},
\end{equation}
where $\kappa$ is a scale parameter. 
In contrast to activation, deactivation is expected to occur at an exponential rate~\cite{gardiner1985handbook}. For simplicity, we assume that the typical switching timescales from the active to inactive state and vice versa are equal, and thus, we set the rate for deactivation to be also $\kappa$.   
The rest of the reactions,   see Fig.~\ref{fig:fig1}, transcription of mRNA and translation of proteins, and  degradation of mRNA and proteins,  are modeled as first-order, exponential processes with rates $a$, $b \gamma$, $\gamma$ and 1,  respectively, where  time is measured in units of inverse protein decay rate. 
The associated WTDs for these five reactions (except binding) are $\psi_{j}(\tau) = \lambda_j \exp(\lambda_j \tau)$,
with $\{\lambda_j\}_{j=1}^5 = \{\ell \kappa, \ell a, n, \gamma m , b \gamma m\}$, where $m$ and $n$ are the mRNA and protein numbers,  and $\ell = \{0, 1\}$ is the promoter's state ($0$ inactive,  $1$ active). 
\begin{figure}[t]
    \includegraphics[width=0.47\textwidth,clip=]{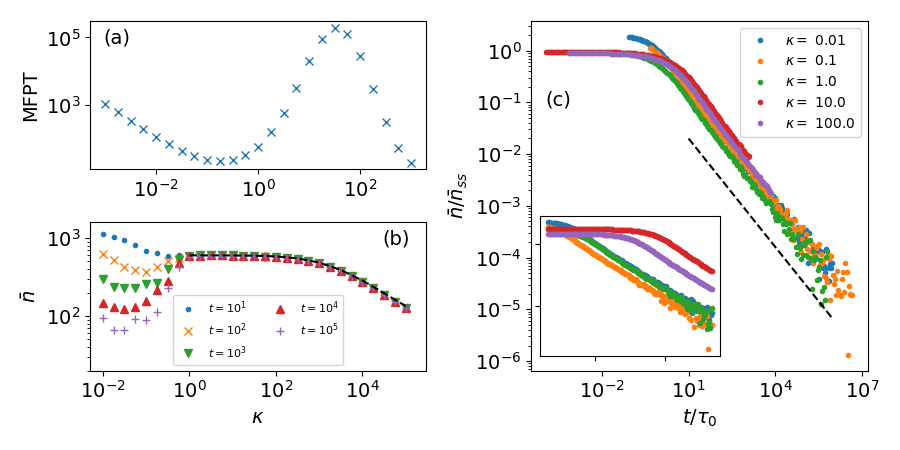}
    \vspace{-5mm}
        \caption{Two-state model. (a) The MFPT versus $\kappa$ for $\alpha = 0.4$.  (b) The mean protein number $\bar{n}$, averaged over $10^3$ simulations,  versus $\kappa$ at different times (legends) for $\alpha = 0.8$. The dashed line is a steady-state solution of Eq.~\eqref{mean_field_full_model} for $\kappa \geq 1$. The theoretical prediction for $\kappa < 1$ is nonstationary and quickly decays. (c) Normalized mean protein copy number $\bar{n}/\bar{n}_{ss}$, averaged over $10^5$ simulations, versus normalized time (see text) for $\alpha = 0.3$. Inset shows the non-normalized averages. The rest of the parameters are $b=1.2$, and  in (a) and (c) $a = 100$ and $\gamma = 10$, while in (b) $a = 1000$, and $\gamma = 100$. }
    \label{fig:fig5}
    \vspace{-5mm}
\end{figure}

As transcription can occur only when the promoter is active and translation is mRNA-dependent, this model is \textit{qualitatively} similar to the SRG model where we modeled the waiting times directly in the protein production. 
In Appendix~\ref{appendixC} we derive the master equation for this set of reactions, and obtain \textit{quantitatively} similar results to the SRG model.  
Here, for $\alpha > 1$, all states including the ground state at $n = m = \ell = 0$, do not exhibit memory at $t \gg 1$, and the dynamics are stationary for all $t \gg 1$. In addition, for $\alpha < 1$ and times $1 \ll t \ll \tau_{typ}$, i.e., longer than the relaxation time but shorter than the typical time $\tau_{typ}$ to sample the ground state, the dynamics are still effectively stationary. In these cases 
the equations for the mean protein and mRNA numbers, $\bar{n}$ and $\bar{m}$, read
\begin{equation} \label{mean_field_full_model}
     \dot{\bar{m}} = a M_{\bar{n}, \bar{m}}^{(0)}/(M_{\bar{n}, \bar{m}}^{(0)} + \kappa) - \gamma \bar{m} \;, \;\;\; \dot{\bar{n}} = \gamma b \bar{m} - \bar{n},
\end{equation}
where $M_{n, m}^{(0)}$ is the leading order of the memory kernel for any state but the ground state, see Appendix~\ref{appendixC}.

In contrast, for $\alpha < 1$ and $t > \tau_{typ}$ the system eventually reaches the ground state $n = m = \ell = 0$ and the dynamics is no longer  stationary. 
Here,  
$
    \bar{n} \sim \bar{m} \sim t^{-1+\alpha}, 
$
see Appendix~\ref{appendixD}, similarly to the SRG model. 
Yet,  to determine the time to reach the ground state in the two-state model, one has to distinguish between two cases: moderate to fast switches $\kappa \geq 1$ and slow switches $\kappa \ll 1$. For $\kappa \geq 1$ we find the same effect as in the SRG model, i.e., the typical time to decay is proportional to the MFPT to the ground state. Here, $\tau_{0} \gg 1$ is governed by the long-lived metastable dynamics and is typically exponential with the mean protein number. Notably, $\tau_{0}$ can be computed using MC simulations, see Fig.~\ref{fig:fig5}(a); it asymptotically decreases as $\kappa \to \infty$. 
In contrast, for $\kappa \ll 1$ the typical time to reach the ground state is no longer governed by the metastable dynamics, since the promoter can be inactive for significantly longer periods than the typical relaxation time of $\OO(1)$, see Fig.~\ref{fig:fig5}(a). This leads to the dynamics reaching the zero state after $\tau_{typ} = \OO(\kappa^{-1})$, resulting in a relatively quick decay $\bar{n} \sim t^{-1+\alpha}$. 

In Fig.~\ref{fig:fig5}(b) we compare numerical solutions of Eq.~\eqref{mean_field_full_model} to simulations, showing stationary dynamics at times $1 \ll t \ll \tau_{typ}$ for $\kappa \geq 1$, and non-stationary dynamics when $\kappa < 1$ for any $t$.
In Fig.~\ref{fig:fig5}(c) our results agree well with simulations for both $1 \ll t \ll \tau_{typ}$ and $t \gg \tau_{typ}$. Here, as in Fig.~\ref{fig:fig4} we normalize $\bar{n}$ by its steady state value found by solving Eq.~\eqref{mean_field_full_model}, and normalize time by the MFPT to reach $n = 0$,  independently obtained from simulations (compare to inset). The collapse occurs at $t\gg \tau_{typ}$  for all values of $\kappa$, and at $t\ll \tau_{typ}$ for $\kappa \geq 1$.  

In summary, we have studied two gene-expression models with delayed protein production due to fat-tailed WTDs. For distributions with diverging mean ($\alpha < 1$) the mean protein number starts to decay after a typical time, which scales as the MFPT to a ground state. Here, the dynamics are ergodic at short times but become nonergodic and display ageing as the ground state is sampled, from which the system can only exit via a non-exponential reaction. 
We also showed that long-range memory may increase  the Fano factor (variance over the mean of the protein distribution) as $\alpha$ decreases, which also holds for a truncated WTD. This effect may be experimentally measurable, provided that the power-law exponent can be adjusted  (e.g., by modifying the molecule's binding affinity to the binding site). 

Although we have focused on gene expression, fat-tailed WTDs may be key in various other fields. For instance, in predator-prey models in movement ecology, diffusion-limited predation or nonergodic foraging may be indicative of power-law waiting times~\cite{vilk2022ergodicity}. Furthermore, long waiting times  also appear in epidemic dynamics, where a large variability in infection and/or recovery periods may lead to  markedly different dynamics~\cite{angstmann2016fractional}.
Our findings may provide valuable insight into these and other dynamical models where long waiting times appear.

OV and MA were supported by the Israel
Science Foundation Grant No. 531/20.


\appendix





\section{Truncated power law} \label{appendixA}
In experimental systems measured power laws are often truncated. In the main text, Fig.~\ref{fig:fig4}(b), we plotted the Fano factor for truncated power laws. The WTD that replaces the one in Eq. \eqref{psi1} is defined as follows 
\begin{equation} \label{psi_truncated}
     \psiu(\tau) = \frac{e^{-\tau/\tau_{\text{cutoff}}}}{(1 + K \tau/\alpha)^{\alpha}}  \left( \frac{1}{\tau_{\text{cutoff}}} + \frac{K}{1 + K \tau/\alpha} \right) . 
\end{equation} 
As shown in Fig.~\ref{fig:fig4}(b), as long as $\tau_{\text{cutoff}}\!>\! 1/K$ (the typical degradation timescale), the Fano factor increases as $\alpha$ decreases.

\section{Ergodicity Breaking} \label{appendixB}
\begin{figure*}[t]
    \includegraphics[width=1.0\textwidth,clip=]{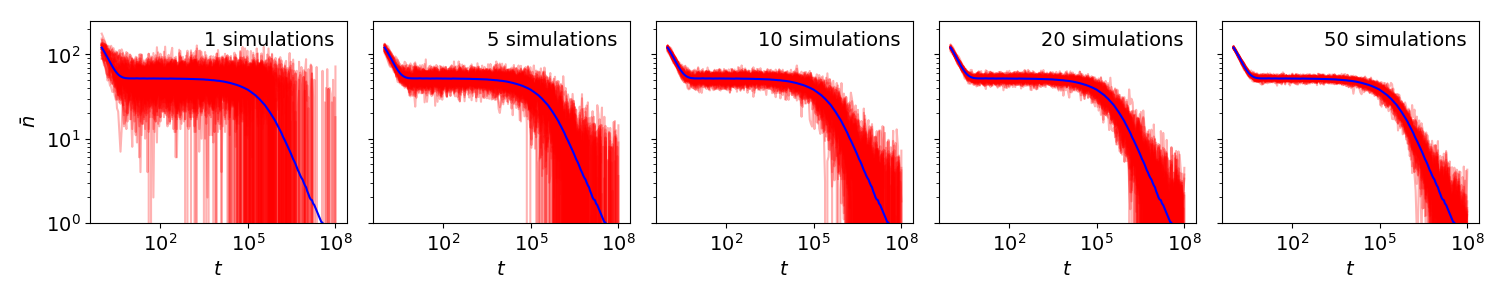}
    \vspace{-9mm}
        \caption{Protein number in the two-state model for $\alpha = 0.3$, $a = 100$, $b=1.2$, and $\gamma = 10$. The blue solid line is the mean protein number, averaged over $10^5$ simulations. In red there are 50 lines, where each represents a trajectory averaged over a varying number of simulations, ranging from no averaging (leftmost panel) to averaging over 50 simulations (rightmost panel). }
    \label{SM:fig1}
\end{figure*}

Ergodicity breaking is formally defined as a disparity between the mean squared displacement (MSD) and time-averaged mean squared displacement (TAMSD). The MSD is defined as the squared displacement of the protein number with respect to a reference number, averaged over an \textit{ensemble} of  independent simulations. 
The TAMSD is given by averaging over the squared displacement of the protein number performed in a time lag $\Delta$~\cite{barkai2012single, metzler2014anomalous}, 
\begin{equation} \label{TASD}
\tasd = \frac{1}{t - \Delta}\int_0^{t - \Delta} [n(t' + \Delta) - n(t')]^2 dt',
\end{equation}
where in this expression an overline denotes time averaging. Note that the same disparity occurs also between other ensemble-averaged and time-averaged observables, and not only the TAMSD~\cite{burov2010aging}. For a Brownian process and $\Delta \ll t$ one obtains $\tasd \sim \Delta \sim \left< x^2(\Delta)\right>$. 
In contrast, if the TAMSD and MSD scale differently, the underlying process is, by definition, nonergodic; that is, the \textit{ensemble} averaging is different from the \textit{time} averaging~\cite{metzler2000random}. 
In many cases, and especially for nonergodic dynamics, it is convenient to compute  the so-called \textit{averaged} TAMSD defined as  
\begin{equation} \label{averagedtasd_def}
\average{\tasd} = 1/N \sum_{i = 1}^{N} \tasd,
\end{equation}
where angular brackets denote ensemble averaging over $N$ simulations (i.e., independent realizations of the protein number). Here, averaging is necessary due to the irreproducible nature of the process (i.e., large diversity across simulations). 
In Fig.~\ref{SM:fig1} we show an example of the large diversity between simulations when the process becomes nonergodic. In the leftmost panel, we plot 50 random simulations compared to the mean protein number, while in the other four panels, each of the 50 lines denotes an average over 5, 10, 20, and 50 simulations. One can see that the variability around the nonergodic phase is larger and more immune to averages over a small number of simulations.

To derive the ergodicity breaking (EB) parameter we define the variability of the TAMSD (the spread of individual TAMSDs around their average) in terms of the dimensionless parameter $ \xi = \tasd/\average{\tasd}$. For many processes, at long measurement times the distribution of $\xi $ satisfies a Mittag-Leffler distribution~\citep{he2008random}
\begin{equation} \label{phi_xi}
\phi(\xi) = \frac{\Gamma^{1/\alpha}(1 + \alpha)}{\alpha \xi^{1 + 1/\alpha}} l_\alpha\left( \frac{\Gamma^{1/\alpha}(1 + \alpha)}{\xi^{1/\alpha}}\right).
\end{equation}
Here, $l_\alpha$ is the one-sided L\'evy stable distribution with the Laplace transform $\mathcal{L}\{l_\alpha (t)\} = \exp(-u^\alpha)$, while $\Gamma(\cdot)$ is the Gamma function. 
For Brownian diffusion, $\alpha \to 1$, $\phi(\xi) \sim \delta(\xi - 1)$, i.e., a sharply peaked distribution around 1. However, for general $\alpha < 1$ the distribution is wide and skewed.
A common measure of ergodicity breaking, which we use in the main text, is the so-called EB parameter, defined for long simulation times $t$ as 
\begin{equation}
    \text{EB} = \average{\xi^2} - \average{\xi}^2.
\end{equation}
In prototypical systems, e.g., CTRW with a power-law WTD, this parameter varies between $\text{EB} = 0$ for an ergodic system to $\text{EB}>0$ for a nonergodic system, where for $\alpha \to 0$ we expect the $\text{EB}$ parameter to be ${\cal O}(1)$~\cite{metzler2014anomalous}.

\section{Derivation of the master equation for the two-state model} \label{appendixC}
To derive the CME for the set of reactions in the two-state promoter model, we repeat the steps detailed for the SRG model based on~\cite{aquino2017chemical}, noting that now we have 6 rates and not only 2. We start by writing the probability density for reaction $i$ to occur at time $t$ while no other reaction $j\neq i$ occurs until $t$:
\begin{equation}
    \phi_i(n, m, \ell, t) = \psi_i(n, m, \ell, t) \prod_{j = 0, j \neq i}^5 \int_{t}^{\infty}\psi_j(n, m, \ell, \tau) d\tau
\end{equation}
A chemical master equation for the probability of having $n$ proteins and $m$ mRNAs at time $t$ when the DNA is in state $\ell$, $P_{n, m}^{(\ell)}(t)$, is then given by 
\begin{eqnarray} \label{Pnm}
    \pd{P_{n, m}^{(\ell)}}{t} = && (-1)^{\ell} \left[\kappa P_{n, m}^{(1)}(t)\! -\! \int_0^t M_{n, m}(t - \tau) P_{n, m}^{(0)}(\tau) d\tau \right] \nonumber \\ && + \bm{A} \Pnml{t} . 
\end{eqnarray}
The memory kernel $M_{n, m}(t)$ is defined in terms of its Laplace transform
\begin{equation} \label{Mkernel_FullModel}
   \tilde{M}_{n, m}(s)  = \frac{s \tilde{\phi}_0(n, m, 0, s)}{1 - \sum_{j=0}^{5} \tilde{\phi}_j(n, m, 0, s)}, 
\end{equation}
and the operator $\bm{A}$ is defined in terms of the step operators $E_k^j f(k) = f(k + j)$ by $\bm{A} = (E_n^1 -1)n +\gamma(E_m^1 -1)m  +\gamma b m(E_n^{-1} -1) + a \ell (E_m^{-1} -1) $. Note that the memory kernel associated with any of the exponential rates $\{\lambda_j\}_{j =1}^5$ can also be calculated in a similar way to Eq.~\eqref{Mkernel_FullModel} and reduces to $M_j(t) = \lambda_j\delta(t)$, giving the form of the operator $\bm{A}$ in Refs.~\cite{gardiner1985handbook, aquino2017chemical}. 
Explicit calculation of the memory kernel in the Laplace variable, \eqref{Mkernel_FullModel}, yields, after some algebra 
\begin{equation}
    \tilde{M}_{n, m}(s) = \frac{(s+\lambda _{tot}) \alpha   E_{\alpha +1}\left(\frac{\alpha  \left(s+\lambda_{tot}\right)}{\kappa }\right)}{e^{-\frac{\alpha  \left(s+\lambda _{tot}\right)}{\kappa }}-\alpha  E_{\alpha +1}\left(\frac{\alpha  \left(s+\lambda
   _{tot}\right)}{\kappa }\right)},
\end{equation}
where $\lambda_{tot} = n + \gamma m + \gamma b m $, and noting that the additional reactions of deactivation and transcription do not contribute when the promoter is in the inactive state. The exponential integral function, $E_{\alpha}(x)$, is defined in the main text.

The memory kernel can be simplified in the limit $s \to 0$ ($t\to \infty$). An immediate result for all states $\lambda_{tot} >0$ is 
\begin{eqnarray} \label{Mnml_lamg0}
    &&\tilde{M}_{n, m}(s) =  M_{n, m}^{(0)} + \OO(s) , \\ && M_{n, m}^{(0)} \equiv \frac{\lambda _{tot} \alpha   E_{\alpha +1}\left(\frac{\alpha  \lambda _{tot}}{\kappa }\right)}{e^{-\frac{\alpha \lambda _{tot}}{\kappa }}-\alpha  E_{\alpha +1}\left(\frac{\alpha  \lambda _{tot}}{\kappa }\right)},\;\;\;\lambda_{tot} > 0 , \nonumber
\end{eqnarray}
which is independent of $s$ in the leading order. Applying the inverse Laplace transform to Eq.~\eqref{Mnml_lamg0} yields $M_{n, m, \ell}(t) \simeq M_{n, m}^{(0)}\delta(t) $ for $\lambda_{tot} > 0$. 
As long as the dynamics are not in the state $n = m = \ell = 0$ we have $\lambda_{tot} > 0$  and this approximation holds. However, for $n = m=\ell = 0$ the only possible reaction is the DNA switching to the active state and we have $\lambda_{tot} = 0$. Here the form of the memory kernel at long times is strongly dependent on $\alpha$:
\begin{eqnarray}    \label{M000}
    \tilde{M}_{0, 0}(s) = &&  \frac{s \alpha   E_{\alpha +1}\left(\frac{\alpha  s}{\kappa }\right)}{e^{-\frac{\alpha  s}{\kappa }}-\alpha  E_{\alpha +1}\left(\frac{\alpha  s}{\kappa }\right)} \\= && \begin{cases}
        \left(\frac{\alpha }{\kappa }\right)^{-\alpha }\frac{ s^{1-\alpha }}{\Gamma (1-\alpha )} + \OO(s^{2-2\alpha}) & \alpha < 1, \\
        \frac{(\alpha -1) \kappa }{\alpha } + \OO(s^\alpha) & \alpha > 1. 
    \end{cases}  \nonumber
\end{eqnarray}
Thus, for $\alpha > 1$ all states do not exhibit memory at $t \gg 1$, suggesting stationary dynamics at long times, similarly to the  exponential WTD case~\cite{shahrezaei2008analytical, assaf2011determining,assaf2013extrinsic}. Moreover, even for $\alpha < 1$ and times $1 \ll t \ll \tau_{typ}$, i.e., longer than the relaxation time but shorter than the typical time $\tau_{typ}$ to sample the ground state $\lambda_{tot} = 0$, the dynamics are expected to be stationary. In addition, when the mean protein number is large the value of the memory kernel for $\lambda_{tot} = 0$  will be negligible. For all of these cases, the steady-state equation for $P_{n, m}^{\ell}$ is given by 
\begin{equation} \label{Pnm_ss}
    0 = (-1)^{\ell} \left[\kappa P_{n, m}^{(1)} - M_{n, m} P_{n, m}^{(0)} \right]  + \bm{A}  P_{n, m}^{(\ell)},
\end{equation}
where the derivation is similar to the one shown above for the SRG. A similar set of equations for a stationary problem was analyzed in Ref.~\cite{assaf2011determining}, and it was shown that the stationary mean-field dynamics of~\eqref{Pnm_ss} follow 
\begin{equation} \label{mean_field_full_model_SI}
    0 \!=\! \dot{\bar{m}} \!=\! a M_{\bar{n}, \bar{m}}^{(0)}/[M_{\bar{n}, \bar{m}}^{(0)} \!+\! \kappa] \!-\! \gamma \bar{m}, \;\;\; 0\!=\! \dot{\bar{n}} \!=\! \gamma b \bar{m} \!-\! \bar{n}.
\end{equation}
As Eqs.~\eqref{mean_field_full_model_SI} are transendental equations for $\bar{n}$ and $\bar{m}$, in general they can only be solved numerically.

\section{Effective two-state model} \label{appendixD}
To show the scaling of the mean number of proteins and mRNA with time, at sufficiently long times, we construct an effective model in which the switches are decoupled from the other cell components. In this effective model the probabilities of the DNA to be in the active and inactive states, $P = \sum_n\sum_m P_{n, m}^{(1)}$ and $Q = \sum_n\sum_m P_{n, m}^{(0)}$, respectively, are given by 
\begin{eqnarray}
    &&\pd{P}{t} = - \kappa P(t) + \int_0^t \psi_0(t - \tau) \kappa P(\tau) d\tau, \label{Pon} \\
    &&\pd{Q}{t} =  \kappa P(t) - \int_0^t \psi_0(t - \tau) \kappa P(\tau)d\tau. \label{Poff}
\end{eqnarray}
The initial condition is $P(0) = 1$, and the probabilities obey $P(t) + Q(t) = 1$ for any $t$. These equations are based on the non-Markovian kinetic rate equations developed in~\cite{kurilovich2020complex}, see also~\cite{doerries2022rate} for further details. Equations~\eqref{Pon} and \eqref{Poff} can be Laplace transformed 
\begin{eqnarray}
    &&s \tilde{P}(s) - 1 = - \kappa \tilde{P}(s) + \kappa \tilde{P}(s)\tilde{\psi}_0(s),\nonumber\\ &&    s \tilde{Q}(s) = \kappa \tilde{P}(s) - \kappa \tilde{P}(s)\tilde{\psi}_0(s).
\end{eqnarray}
Solving for $\tilde{P}$ and $\tilde{Q}$ we find 
\begin{equation} \label{Ponoff_tildes}
    \tilde{P}(s) = \frac{1}{\kappa  (1-\tilde{\psi}_0 (s))+s} ,\;\;\;\;
    \tilde{Q}(s) = \frac{\kappa  (1-\tilde{\psi}_0 (s))}{s [\kappa  (1-\tilde{\psi}_0 (s))+s]}.
\end{equation}
Assuming $t\gg \kappa^{-1}$, i.e., that the total time is much larger than the time of any switch, we then have $s \ll \kappa$ and: 
\begin{equation} \label{tilde_psi_approx}
    \tilde{\psi}_0(s) = 
    \begin{cases}
        1- \frac{\alpha  s}{\kappa (\alpha -1) } + \OO(s^\alpha) & \alpha >1,  \\
         1-\Gamma (1-\alpha ) \left(\frac{\alpha  s}{\kappa }\right)^{\alpha } + \OO(s) & \alpha < 1. 
    \end{cases}
\end{equation}
For $\alpha < 1$, we substitute Eq.~\eqref{tilde_psi_approx} into Eq.~\eqref{Ponoff_tildes} and perform the inverse Laplace transform. This yields
\begin{equation} \label{Pon_sol}
    P \simeq \frac{\sin (\pi  \alpha ) }{\pi \alpha }\left(\frac{\kappa  t}{\alpha }\right)^{-(1-\alpha )},\quad\quad Q=1-P.
\end{equation}
Here, the probability $P$ is slowly decaying with time. 
To find the average mRNA and protein numbers we assume that due to the slow DNA dynamics the equations for the mRNA and proteins depend only on $P$:
\begin{equation} \label{mrna_mf}
    d\bar{m}/dt = - \gamma \bar{m}(t) + a P\; , \;\; d\bar{n}/dt = \gamma b \bar{m}(t) - \bar{n}.
\end{equation}
Substituting Eq.~\eqref{Pon_sol} into Eqs.~\eqref{mrna_mf}, solving for $\bar{m}$ and $\bar{n}$ and finally approximating the result for $t \gg 1$ we find 
\begin{equation} \label{mrna_mf_sol}
    \bar{n}(t) \simeq \gamma b \bar{m}(t) \simeq a b P. 
\end{equation}
Note that, this result can also be obtained by assuming that the derivative with respect to time on the left-hand side of both Eqs.~\eqref{mrna_mf} are small with respect to the right-hand side. This directly leads to Eqs.~\eqref{mrna_mf_sol}. 

We note that for $\alpha > 1$ we can also substitute Eq.~\eqref{tilde_psi_approx} into~\eqref{Ponoff_tildes} and perform the inverse Laplace transform. This yields $P \simeq (1-\alpha)/(1-2\alpha)$, i.e., the probability approaches a constant at long times. For $\alpha \gg 1$ this reduces to $P \simeq 1/2$ as expected; yet, for $\alpha \to 1_{+}$ the probability  satisfies $\lim_{\alpha \to 1_{+}} P = 0$. Although this seems counter-intuitive, this result is only valid in the limit of $s \ll \kappa (\alpha-1)$ or $t \gg [\kappa (\alpha-1)]^{-1}$, see Eq.~\eqref{tilde_psi_approx}.

\bibliography{references}

\end{document}